\documentclass[twocolumn]{aastex61}

\usepackage{mathtools}
\usepackage{booktabs}
\usepackage{array}
\usepackage{graphicx}
\usepackage{color}

\newcolumntype{C}[1]{>{\centering\arraybackslash}p{#1}}

\received{September 4, 2018}
\revised{February 4, 2019}
\accepted{February 27, 2019}

\submitjournal{ApJ}

\shorttitle{Size-scaling of clumps}
\shortauthors{Ali et al.}

\begin{document}

\title{Size-scaling of clump instabilities in turbulent, feedback regulated disks}

\correspondingauthor{Kamran Ali}
\email{kamran.ali@icrar.org}


\author{Kamran Ali}
\affiliation{International Centre for Radio Astronomy Research (ICRAR),
M468, University of Western Australia, 35 Stirling Hwy, Crawley, \\
WA, 6009, Australia}

\author{Danail Obreschkow}
\affiliation{International Centre for Radio Astronomy Research (ICRAR),
M468, University of Western Australia, 35 Stirling Hwy, Crawley, \\
WA, 6009, Australia}

\author{Liang Wang}
\affiliation{International Centre for Radio Astronomy Research (ICRAR),
M468, University of Western Australia, 35 Stirling Hwy, Crawley, \\
WA, 6009, Australia}

\author{Deanne B. Fisher}
\affiliation{Centre for Astrophysics and Supercomputing, Swinburne University of Technology, P.O. Box 218, Hawthorn, VIC 3122, Australia}

\author{Karl Glazebrook}
\affiliation{Centre for Astrophysics and Supercomputing, Swinburne University of Technology, P.O. Box 218, Hawthorn, VIC 3122, Australia}
\author{Ivana Damjanov}
\affiliation{Harvard-Smithsonian Center for Astrophysics, 60 Garden St., Cambridge, MA 02138, USA}
\affiliation{Department of Astronomy \& Physics, Saint Mary's University, 923 Robie Street, Halifax, NS, Canada, B3H 3C3}

\author{Roberto G. Abraham}
\affiliation{Department of Astronomy and Astrophysics, University of Toronto, 50 St. George St., Toronto, ON M5S 3H8, Canada}
\affiliation{Dunlap Institute, University of Toronto, 50 St. George St., Toronto, ON M5S 3H8, Canada}
\author{Emily Wisnioski}
\affiliation{Research School of Astronomy and Astrophysics, Australian National University, Canberra, ACT 2611, Australia}
\affiliation{ARC Centre of Excellence for All Sky Astrophysics in 3 Dimensions (ASTRO 3D)}


%
%



\begin{abstract}
We explore the scaling between the size of star-forming clumps and rotational support in massively star-forming galactic disks. The analysis relies on simulations of a clumpy galaxy at $z=2$ and the observed DYNAMO sample of rare clumpy analogs at $z\approx0.1$ to test a predictive clump size scaling proposed by \citet{Fisher2017ApJ...839L...5F} in the context of the Violent Disk Instability (VDI) theory. We here determine the clump sizes using a recently presented 2-point estimator, which is robust against resolution/noise effects, hierarchical clump substructure, clump-clump overlap and other galactic substructure. After verifying Fisher's clump scaling relation for the DYNAMO observations, we explore whether this relation remains characteristic of the VDI theory, even if realistic physical processes, such as local asymetries and stellar feedback, are included in the model. To this end, we rely on hydrodynamic zoom-simulations of a Milky Way-mass galaxy with four different feedback prescriptions. We find that, during its marginally stable epoch at $z=2$, this mock galaxy falls on the clump scaling relation, although its position on this relation depends on the feedback model. This finding implies that Toomre-like stability considerations approximately apply to large ($\sim\rm kpc$) instabilities in marginally stable turbulent disks, irrespective of the feedback model, but also emphasizes that the global clump distribution of a turbulent disk depends strongly on feedback.
\end{abstract}


\section{Introduction} \label{sec:intro}

High redshift ($z>1$) star-forming galaxies show a more irregular and clumpy structure than local spiral galaxies \citep{Elmegreen2004ApJ...604L..21E, Elmegreen2006ApJ...650..644E}. The luminous clumps measure up to $1\rm~kpc$ in radius \citep{Swinbank2012ApJ...760..130S, Fisher2017MNRAS.464..491F}. They are sites of extreme star formation that collectively reach star formation rates (SFRs) up to $100\ M_{\odot} \rm{yr^{-1}}$ \citep{Genzel2006Natur.442..786G,  Stark2008Natur.455..775S}. In contrast, local main sequence galaxies, such as the Milky Way, only support SFRs of a few $M_{\odot} \rm{yr^{-1}}$ \citep{Licquia2015ApJ...806...96L} and this star formation is hosted in much smaller `Giant' Molecular Clouds (GMCs), measuring less than $100\rm pc$ in radius \citep{Bolatto2008ApJ...686..948B}. This difference between high-redshift and local star-forming systems parallels the strong evolution of the global comoving SFR density, which peaked at $z\sim2-3$ \citep{Hopkins2006ApJ...651..142H, Yuksel2008ApJ...683L...5Y} and has since declined by more than an order of magnitude. Hence, understanding the origin and physics of massive star-forming clumps is an important jigsaw piece in modelling the evolution of galaxies.

The most common scenarios for the formation of star-forming clumps can be grouped into in-situ and ex-situ processes. Ex-situ clump formation relates to environmental interactions such as star-bursting major mergers and minor mergers, where the merging satellite becomes a clump of its new host galaxy. 

In-situ clump formation normally invokes the theory of Violent Disk Instabilities (VDIs), in which a turbulent, rotating disk fragments into gravitationally bound sub-structures \citep{Bournaud2009}. Due to high velocity dispersion, the Jeans' scales, the length at which thermal expansion and contraction due to gravity are in equilibrium, can reach up to $1~\rm kpc$, only a few times shorter than the characteristic scale of the entire disk. Such large Jeans' lengths are a necessary, but insufficient condition for large clumps to form. It is also required that instabilities of this size are \textit{not} stabilised by shear forces -- a non-trivial requirement in rotating disks \citep{Burkert2010}. A metric to quantify these instabilities is the Toomre parameter $Q$ \citep{Toomre1964ApJ...139.1217T} which measures the ratio between the outward pressure (thermal$+$dynamical) and gravitational force within a gas cloud. In the approximation of an axially symmetric disk, the situation of marginal stability can be expressed as $Q\approx1$, where $Q$ is a two-component (gas+stars) extension \citep[e.g.][]{Romeo2011} of the Toomre stability parameter. Using this ansatz several studies found that the marginal stability of clumpy disks can be attributed to high gas fractions \citep{Dekel2009ApJ...703..785D,Genzel2011ApJ...733..101G,Fisher2014,Wisnioski2015ApJ...799..209W, White2017ApJ...846...35W} and/or low angular momentum, with the latter being likely the dominant cause \citep{Obreschkow2015}. However, this $Q$-based ansatz remains debated and may require the inclusion of additional non-linear processes \citep{Inoue2016}.

Distinguishing between different clump formation scenarios is not trivial from an observational viewpoint (e.g.~\citealp{Glazebrook2013PASA...30...56G}). A purely morphological analysis of the CANDELS data \citep{Guo2015ApJ...800...39G} suggests that the incidence of clumps in massive ($M_{\rm *}>10^{10}M_{\odot}$) star-forming galaxies at $z\approx0.5-3$ is consistent with the VDI model, whereas minor mergers might be important for lower mass galaxies and at lower redshifts. Using HST images of galaxies with spectroscopic redshifts from the VIMOS Ultra Deep Survey (VUDS), \citet{Ribeiro2017A&A...608A..16R} analyse the number and luminosity statistics of clumps in individual galaxies at $2 \lesssim z \lesssim 6$ and again conclude that VDIs are probably the dominant cause of clump formation, rather than mergers. Additional circumstantial support for VDIs as the dominant origin of massive clumps comes from resolved kinematic studies (e.g.~\citealp{Tacconi2013ApJ...768...74T}) revealing that the majority of high-redshift star-forming galaxies are rotationally supported disks. However, \citet{Forster2009ApJ...706.1364F} and \citet{Law2009ApJ...697.2057L} find that it is possible for galaxies undergoing strong mergers to display a rotation profile that closely resembles that of a rotating disk.

Of course, resolved imaging and spectroscopy of individual clumps would enable much more stringent tests, however, this is normally hampered by instrumental limitations -- even at HST resolution sub-kpc scales at $z=2$ are barely resolved. To beat this limitation \citet{Dessauges-Zavadsky2018MNRAS.479L.118D} analyzed a sample of strongly lensed galaxies. They found that the mass function of clumps follows a power law of slope~$-2$ which is consistent with clumps forming in-situ by turbulent fragmentation. However, since the magnification of strong lensing is model dependent and acts only in a single direction, the interpretation of such data remains difficult.

An alternative approach to studying high redshift galaxies consists of using their lower redshift `analogs'. This is the leading idea of the DYNAMO sample, detailed in Section~\ref{subsec:dynamo}. Relying on a $Q$-based approximation, \citet{Fisher2017ApJ...839L...5F} (hereafter F17) predicted and observationally confirmed that clumps formed in-situ obey a scaling relation between the clump radius and the velocity dispersion (and, by extension, the gas fraction) of their parent disk (see also \citealp{Wisnioski2012,Livermore2012MNRAS.427..688L}). Using this relation, F17 explicitly showed that expectations from a minor-merger scenario are not likely to form most clumps in DYNAMO galaxies. Hence, this relation is a promising way to distinguish between clump formation scenarios, as well as to probe the inner physics of these heavily star-forming objects. 


The use of scaling relations to test ideas for the origin of clumps raises important challenges:
\begin{itemize}
\item \textit{Clump size measurement:} The definition and measurement of the characteristic clump size should be robust against (1) variations in observational resolution/noise, (2) the complex hierarchical substructure of clumps \citep{Elmegreen2011EAS....51...31E}, (3) the potential random overlap of clumps, and (4) the presence of other, similarly-sized galactic substructures, such as bars and spiral arms.
\item \textit{Physics of scaling relation:} The model of F17, relating characteristic clump size to the disk's velocity dispersion, relies on a simplification of the Toomre stability criterion \citep{Toomre1964ApJ...139.1217T}, which makes a number of approximations and bypasses the possible strong feedback-regulation \citep{Genel2012} within star-forming clumps.
\end{itemize}

Here, we address these challenges using an advanced statistical method applied to both observations and simulations of clumpy galaxies. We use the well-tested 2-point statistics-based method of \citet{Ali2017ApJ...845...37A} (summarized in Section \ref{sec:background}) to measure the `characteristic' clump size in resolved images of star-formation rate traces. In Section \ref{sec:observations}, this method is applied the full set of 10 nearby clumpy galaxies from the DYNAMO (DYnamics of Newly-Assembled Massive Objects) survey \citep{Green2014MNRAS.437.1070G} that have been followed up by the Hubble Space Telescope (HST) in H$\alpha$ and continuum emission by \citet{Fisher2017MNRAS.464..491F}. We show that the VDI scaling relation of F17 holds for the clumpy galaxies (except for `new' mergers) when analyzed in this way. In Section \ref{sec:simulations}, we use four realizations of a simulated control galaxy with four different stellar feedback modes in order to (1) verify the scaling relation of F17 in a more realistic model and (2) check if this scaling relation applies irrespective of the feedback model. Section \ref{sec:discussion} gives a synthesis of the results and brief conclusion.

\section{Background: 2-point clump scale} \label{sec:background}

This section summarizes the statistical estimator of the characteristic clump size introduced by \cite{Ali2017ApJ...845...37A} (hereafter A17). The interested reader is referred to that paper for details beyond the brief summary presented here.

In A17, we found that the characteristic scale $r_{\rm clump}$ of the clumps in a star-formation density map is related to the maximum point $r_{\rm peak}$ of the weighted two-point correlation function (w2PF)
\begin{equation} \label{eq:w2pf}
r^{\gamma}\xi_{2}(r)\ \forall\ \gamma>0,
\end{equation}
where $r$, $\xi_{2}(r)$ and $\gamma$ are the length scale, the two-point correlation function of the map and a positive exponent, respectively. For randomly positioned clumps with circular 2D-Gaussian density profiles of standard deviation $r_{\rm clump}$, the exact analytical expectation is
\begin{equation} \label{eq:clump=peak}
r_{\rm clump} = \frac{r_{\rm peak}}{\sqrt{2\gamma}}.
\end{equation}

\begin{figure*}
\includegraphics[width=1\textwidth]{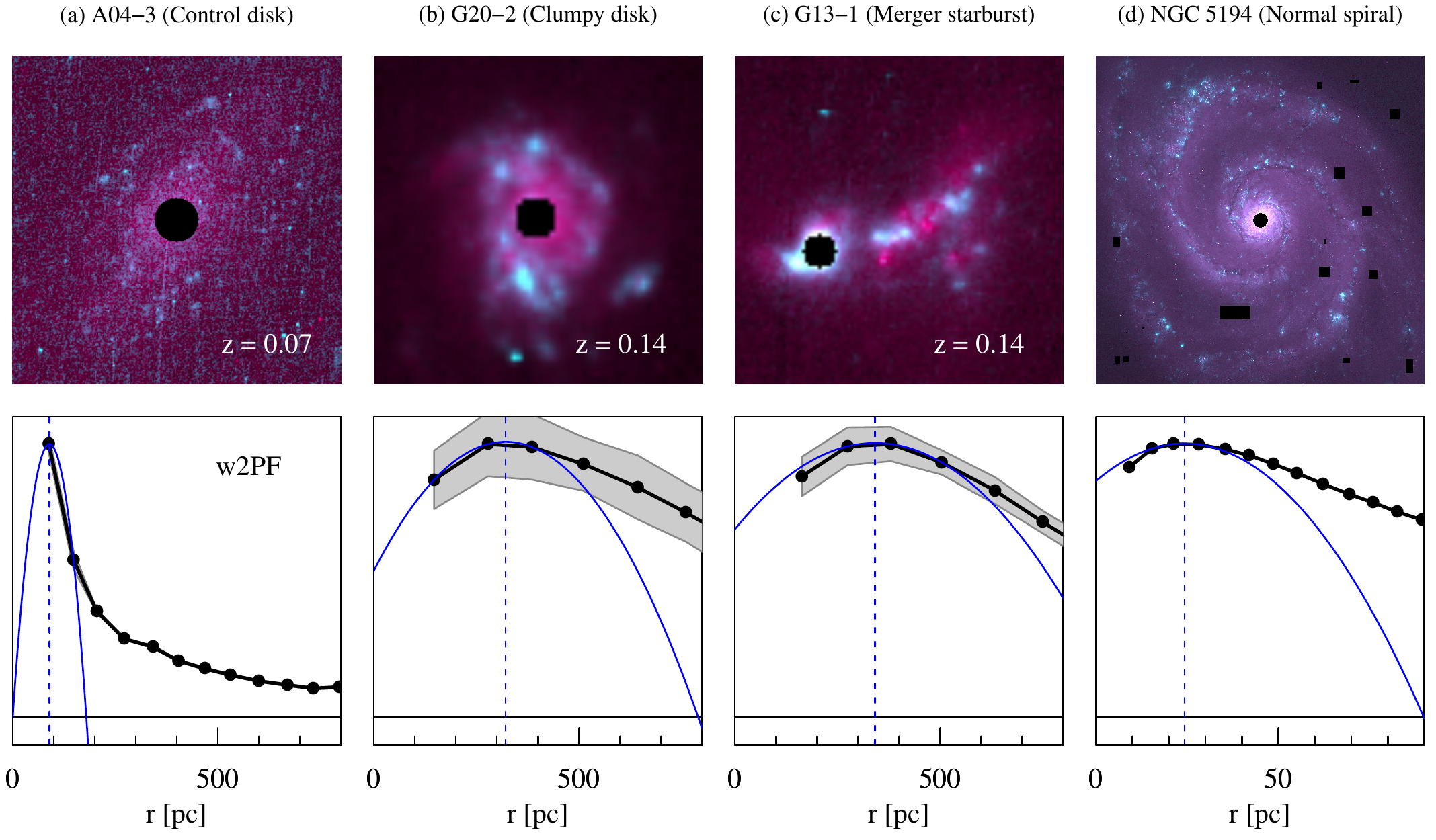}
\caption{Example of four galaxies in the observed sample: (a) the `normal' control galaxy in the DYNAMO-HST sample with regular GMCs that cannot be resolved, (b) one of 7 clumpy DYNAMO-HST galaxies with rotation supported disks, (c) one of two merging starbursts in DYNAMO-HST, (d) NGC 5194, a local ($z\approx0$) spiral galaxy with GMCs. The H$\alpha$ and continuum maps are displayed in cyan and red, respectively. The bottom row shows their w2PF (for $\gamma=1/2$), in arbitrary units on the y-axis, 1$\sigma$ uncertainty (shaded region) and their $r_{\rm clump}=r_{\rm peak}$ value (dashed line), determined by fitting a parabola (blue) at around the maxima. A distance of  $8\ \rm Mpc$  \citep{Karachentsev2004AJ....127.2031K} is used in our values for NGC 5194, but note that $r_{\rm clump}/r_{\rm disk}$ is distance-independent. }
\label{fig:dynamo_galaxies}
\end{figure*}

In particular, if $\gamma=1/2$, then $r_{\rm clump}=r_{\rm peak}$. Real clumps are neither Gaussians of identical size, nor are they randomly positioned across the galaxy. However, the w2PF method is robust against these deviations, as shown in A17 for the following reasons.

Firstly, star-forming clumps normally exhibit a size distribution (often following power law between number and size, see \citealp{Oey1998AJ....115.1543O}) and they exhibit a complex, hierarchical substructure down to the scale of individual star-forming `cores'. Using mock images of clumps with an overall Gaussian density, but fractal substructure, drawn from realistic size-distributions, we showed numerically that the w2PF method recovers the mass-weighted clump size of the input model within 20\%, irrespective of the precise size-distribution and fractal substructure. Furthermore, the w2PF is robust against changing resolution, as long as the Point Spread Function (PSF) is smaller than the mass-weighted clump size, and also robust to different types of noise (with white, blue, red spectra) up to an RMS pixel-noise as high as the integrated flux of the brightest clumps (Fig.~3 in A17).

Secondly, the clump positions in real galaxies are not random, but they follow the global density structure of the disk, such as a roughly exponentially decreasing surface density with spiral arms and rings. We found that these galactic structures impact the clump size measurement via the w2PF, but can be removed in the same way that window functions and selection functions can be removed when measuring the 2PCF of cosmic large-scale structure. That is, in expression (\ref{eq:w2pf}), we must define $\xi_{2}(r)$ as the 2PCF in excess of galactic substructure other than clumps. This can be done, for instance, using the classic Landy-Szalay estimator \citep{LS1993} 
\begin{equation} \label{eq:LS}
\hat{\xi}_{\rm LS}(r)= \frac{DD(r)-2DR(r)+RR(r)}{RR(r)},
\end{equation}
where the fields $D$ and $R$ are 2D fields. The $D$ contains the clumpy structure which we wish to measure and extra spurious galactic structure while the $R$ field consists of only the galactic structure wish we wish to remove. Hence in measuring clump sizes via 2PCF estimator it is vital we select an $R$-field which masks the excess correlation. The functions $DD$, $DR$ and $RR$ are defined as
\begin{equation} \label{eq:normalise}
\begin{split}
XY(r)\equiv\frac{1}{\sum X\sum Y} \sum_{\rm |\pmb{r_1-r_2}|\in(r\pm\Delta r/2)} X(\pmb{r_1})Y(\pmb{r_2}).
\end{split}
\end{equation}
The parameter $\Delta r$ is the bin width of the regularly distributed scale lengths $r$. Since Eq.~(\ref{eq:LS}) effectively removes correlations present in the $R$-field from the $D$-field we take a map of the older stellar population as $R$-field while that of the newly formed stars as $D$-field. For the DYNAMO-HST sample we use H$\alpha$ map as the $D$-field and the continuum map by using it as $R$-field. In the case of the simulations we use stars formed within $10\ \rm Myr$, which corresponds to the lifetime of O-stars, as $D$-field while taking the whole stellar population as the $R$-field. This removes spurious correlation added to the 2PCF by the galaxy disk structure.

Finally, we must choose a value of $\gamma$ when computing the w2PF (Eq.~\ref{eq:w2pf}). For a hypothetical infinitely extended field of Gaussian clumps, any positive value will result in an accurate estimation of the clump scale $r_{\rm clump}$ (via Eq.~\ref{eq:clump=peak}). However, in realistic circumstances, larger values can help suppress spurious small-scale structure, not already removed via the R-field in Eq.~(\ref{eq:LS}), whereas smaller values can suppress spurious large-scale structure. In A17 we adopted the fiducial $\gamma=1/2$, which leads to good results for mock images of galaxies with realistic noise. We here apply this value to all observed galaxies. In the case of our simulated disks, we find that a slightly larger value (we pick $\gamma=1$) allows us to avoid contamination by spurious small-scale structures associated with two-body relaxation present in SPH-based simulations (e.g.~\citealp{Power2016MNRAS.462..474P}).


\section{Clump-scalings in observed galaxies} \label{sec:observations}

In this section we first describe the observational data used in this study. We then apply the w2PF and compare the robustly estimated clump sizes to those measured by F17. Finally, we gauge the degree to which the scaling relationship of F17 holds for the DYNAMO-HST sample.

\subsection{Sample of high-$z$-analogs}\label{subsec:dynamo}

The DYNAMO (DYnamics of Newly-Assembled Massive Objects) galaxies \citet{Green2014MNRAS.437.1070G} were selected from the Sloan Digital Sky Survey (SDSS) \citep{York2000AJ....120.1579Y} as the objects with the most extreme H$\alpha$ luminosities ($L_{\rm H\alpha} > 10^{42}\ \rm erg\ s^{-1}$). Follow-up integral field spectroscopy observations were used to identify a subsample of high-dispersion systems, which excludes AGN. A sub-sample of 9 such galaxies, as well as one control galaxy (A04-3) with normal H$\alpha$ luminosity, were then observed with the Hubble Space Telescope (HST) Advanced Camera for Surveys Wide-field Camera by \citet{Fisher2017MNRAS.464..491F}. All galaxies in this DYNAMO-HST sample, except for the control object, show massive clumps, reminiscent of those seen at higher redshift, when degraded to $z=1$ resolution. Two galaxies in the DYNAMO-HST clearly look like systems about to undergo a major merger, while the other 8 show regular morphologies and rotation-supported disks in H$\alpha$ kinematics.

The HST data used in this paper are H$\alpha$ maps showing newly formed stars and a continuum image showing the older stellar population. The H$\alpha$ emission was observed using the ramp filters FR716N and FR782N within a 2\% bandwidth and integrated for 45 minutes. The continuum maps used the FR647M filter with an integration time of 15 minutes. The final H$\alpha$ image is generated by subtracting the continuum map from the H$\alpha$ map. The complete reduction process is given in \citet{Fisher2017MNRAS.464..491F}.

The sample consists of galaxies which are consistent with clump formation scenarios resembling self-gravity instabilities as well as major mergers. For certain galaxies the resolution is not sufficient to measure the clump size $r_{\rm clump}$. Figure \ref{fig:dynamo_galaxies} shows the variety of galaxies analysed in this study along with their w2PF: (a) the control galaxy of the DYNAMO-HST sample without significant clumps, (b) one of 7 clumpy disk with no signs of mergers in the DYNAMO-HST sample, (c) one of two merging starbursts, and (d) the local spiral galaxy NGC 5194. Since the primary goal of our analysis is to measure the clump scale we mask the central region and foreground stars as shown in Figure \ref{fig:dynamo_galaxies} (black). We then compute the w2PF and fit a parabola around the maxima to infer $r_{\rm clump}$ at sub-pixel resolution. In the control object (a), the size of the star-forming regions lies below the resolution and hence the w2PF only provides an upper bound.

\begin{figure}
\includegraphics[width=1\columnwidth]{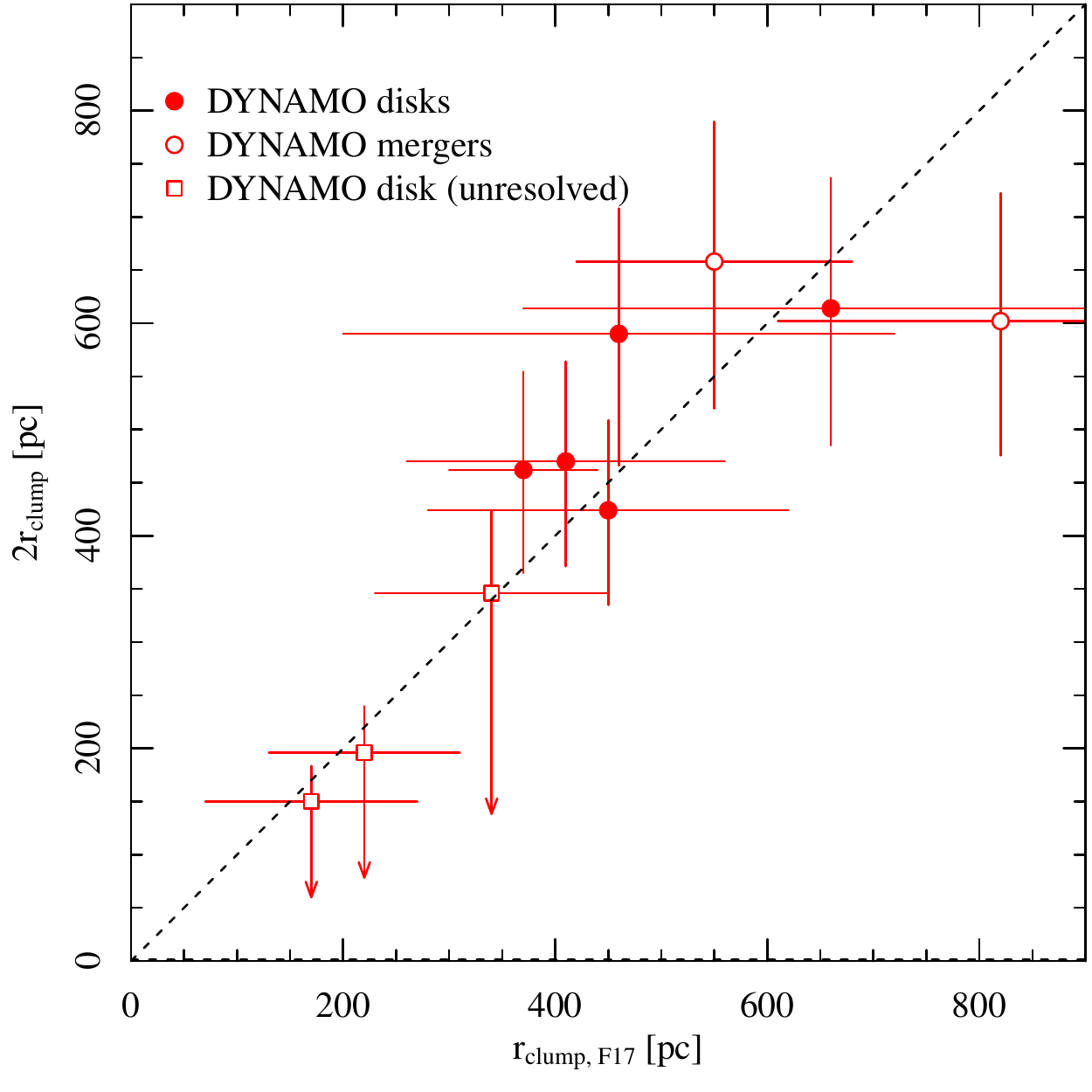}
\caption{Comparison of clump sizes estimated by the w2PF method, $2r_{\rm clump}$, and a previous study by F17 $r_{\rm clump, F17}$. The two estimates are in good agreement as they fall on the one-to-one (dotted) line within their 1-$\sigma$ uncertainties.}
\label{fig:david_compare}
\end{figure}

\subsection{Results}

Comparing \emph{twice} the clump size $r_{\rm clump}$ estimated using the w2PF of A17 with those measured by F17 ($r_{\rm clump, F17}$) we find a good agreement, within 1-$\sigma$ uncertainty, as shown in Figure \ref{fig:david_compare}. Each point in this figure corresponds to a galaxy-average. The method of A17 naturally returns the luminosity-weighted average clump size, which is a converged quantity, even in the presence of hierarchical substructure, given the steep power-law distribution of the substructure (see A17 for details). F17 measure the size of each clump individually and then take the average. This method would result in smaller clump sizes, if individual clumps were resolved into sub-clumps. However, given the current resolution limit, no such substructure is detected. Explicitly, F17 identify the brightest peaks (relative to a smoothing mask) as clumps. They then fit these clumps with a 2D-Gaussian profile and compute an effective radius as the geometric mean of the major and minor half-axes. Finally, they take twice the average of all clumps within the galaxy as $r_{\rm clump, F17}$ due to which we compare their values to $2r_{\rm clump}$.

The errors on $r_{\rm clump}$ are computed by adding in quadrature the uncertainty due to \emph{Sample variance}, \emph{deblurring} and \emph{image noise}. \emph{Sample variance} originates from the fact that we observe only one instance of the galaxy. \emph{Deblurring} is the process of removing the contribution of the PSF from $r_{\rm clump}$ and hence the uncertainty scales in proportion to width of PSF compared to $r_{\rm clump}$. Finally, \emph{image noise} is due to the noise present in H$\alpha$ map. In A17, we explored these three sources of uncertainty in detail and here use the tabulated values to estimate the uncertainty in clump size of each galaxy. The clump sizes have been corrected for the PSF.

\begin{figure}
\includegraphics[width=1\columnwidth]{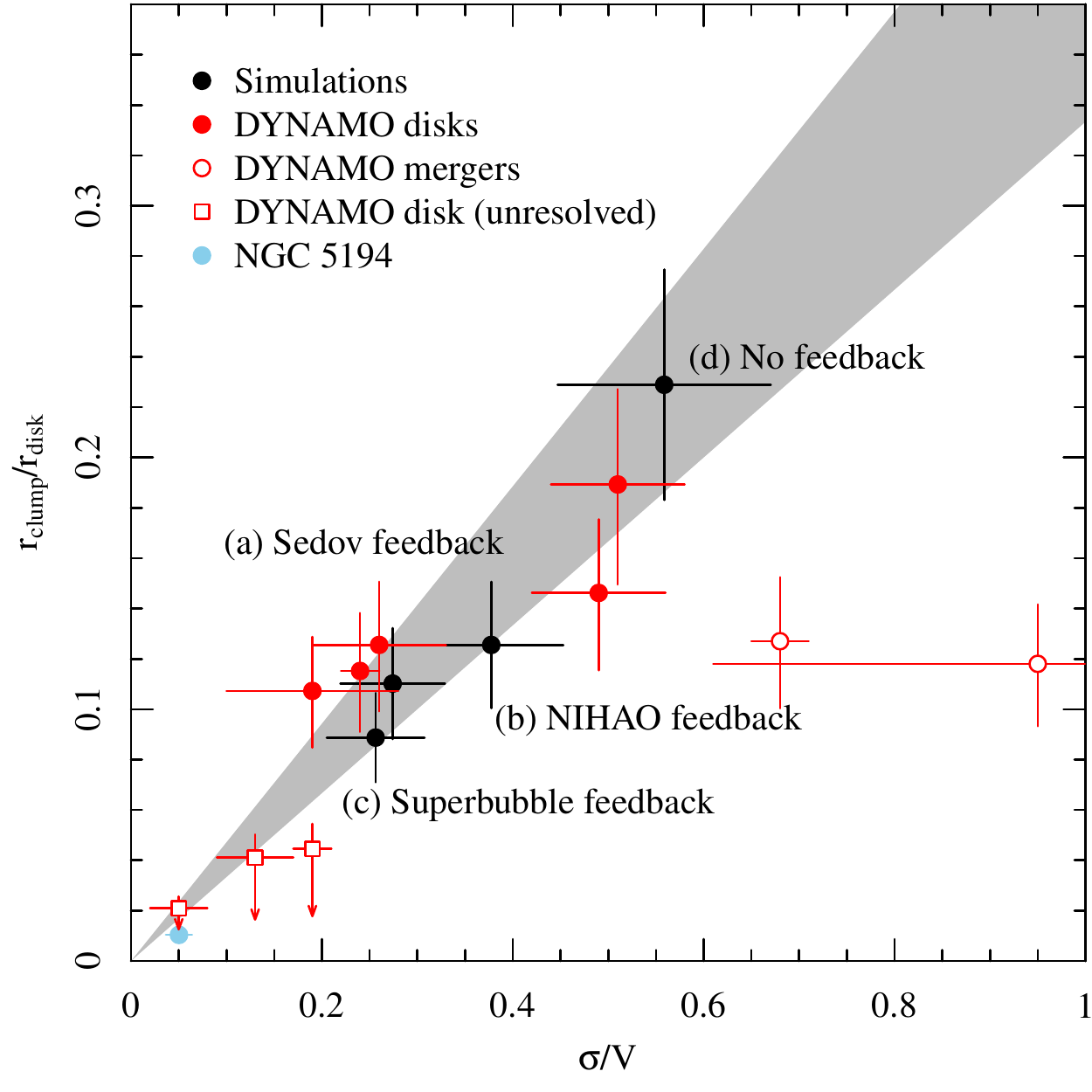}
\caption{Relationship between the clump-to-disk size and velocity dispersion-to-rotation velocity ratios in the DYNAMO-HST sample (red), NGC 5194 (blue) and the simulated galaxy (black). In the case of disk galaxies the average clump properties fall within the maximum and minimum allowed F17 scaling relation (shaded region) while the merging systems deviate significantly from this region due to the large Toomre parameter. This agreement is also seen in simulations regardless of the feedback model.}
\label{fig:all_points}
\end{figure}

Figure \ref{fig:all_points} is a reproduction of the F17 Figure 2 (left) and compares the clump sizes measured using the w2PF to the theoretical model of F17. This VDI model assumes a marginally stable disk, in the sense of an average Toomre \citep{Toomre1964ApJ...139.1217T} parameter $Q\approx1$. A simple calculation then results in the prediction that the clump-to-disk scale ratio is proportional to the gas dispersion-to-rotational velocity ratio,
\begin{equation}
	\frac{r_{\rm clump}}{r_{\rm disk}} = a\frac{\sigma}{V},
\end{equation}
where we adopt the definition of A17 that $r_{\rm clump}$ is the Gaussian clump size of eq.~(\ref{eq:clump=peak}) and $r_{\rm disk}$ is the effective radius. (Note that F17 define both values a factor 2 higher, leaving their ratio unchanged.) The proportionality factor $a$ depends on the shape of the rotation curve and is expected to vary between $1/3$ (Keplerian potential) and $\sqrt{2}/3$ (isothermal potential). The allowed range between these two proportionality factors is shown as grey shading in Figure \ref{fig:all_points}. The measurements are consistent with this model, except in the case of the two merging systems (open circles). This confirms the findings of F17 that the clump size scaling relation can help distinguish between major mergers and other scenarios of clump formation.

Figure \ref{fig:all_points} also shows the local main sequence galaxy NGC 5194 (whirlpool galaxy, M51a) in H$\alpha$ and $I$-band maps obtained from the Advanced Camera for
Surveys on board the HST \citep{Mutchler2005AAS...206.1307M}. Using the measurements of $r_{\rm disk}$, $\sigma_{\rm gas}$ and $V$ from \citet{Leroy2008AJ....136.2782L} we notice that NGC 5194 lies below the F17 VDI scaling relation. This is in agreement with \citet{Leroy2008AJ....136.2782L} who find a median value of $Q\approx2-3$ indicating a stable disk (except for in the dense regions of the spiral arms). The new clump size measurements validate the scaling relations presented by F17 as a way of differentiating between clump formation scenarios. 

Open squares in Figure \ref{fig:all_points} denote upper limits for three systems where the clumps are too small for a reliable size determination, in the sense that the peak position of the w2PCF is consistent with the standard deviation of the PSF. The lowest of these points is the control galaxy A04-3 shown in Figure~\ref{fig:dynamo_galaxies}~(a), which by choice does not exhibit large clumps. Two of upper limits seem to lie somewhat below the relation. This might be explained by the fact that these systems are, in fact, rather stable disks ($Q>1$), similarly to NGC 5194.

We also compare the measured clump sizes to the disk thickness estimated in previous studies for three galaxies within our study. The power spectrum, Fourier equivalent of the 2PCF, has been widely used as an indicator of the disk thickness (\citealp{Elmegreen2001ApJ...548..749E}, \citealp{Combes2012A&A...539A..67C}). The fractal nature of the 2D galaxy structure gives a different power law as compared to a 3D structure. As the exponent changes at the transition between 2D and 3D behaviour the correlation function is likely to give a turning point at the associated scale height. We have attempted to mitigate this effect by taking a non-flat $R$-field unlike prior studies, however, the estimated disk thickness changes depending on the wavelength band used to observe the galaxy\citep{Elmegreen2013ApJ...774...86E}. To ensure we are not simply measuring the thickness parameter we compare our $2r_{\rm clump}$ measurements to scale height values determined by previous studies for galaxies G04-1, G20-2 and NGC5194. The clump sizes of $\{590,614,58\}$ (in parsecs) only match one scale height measurement $\{131,562,200\}$ \citep{Bassett2014MNRAS.442.3206B, Pety2013ApJ...779...43P} for the galaxy G20-2 but differ significantly for other two galaxies including the most resolved local galaxy. Hence, we find it unlikely that the w2PF turns over at the scale height of the galaxy.

\begin{figure*}
\includegraphics[width=1\textwidth]{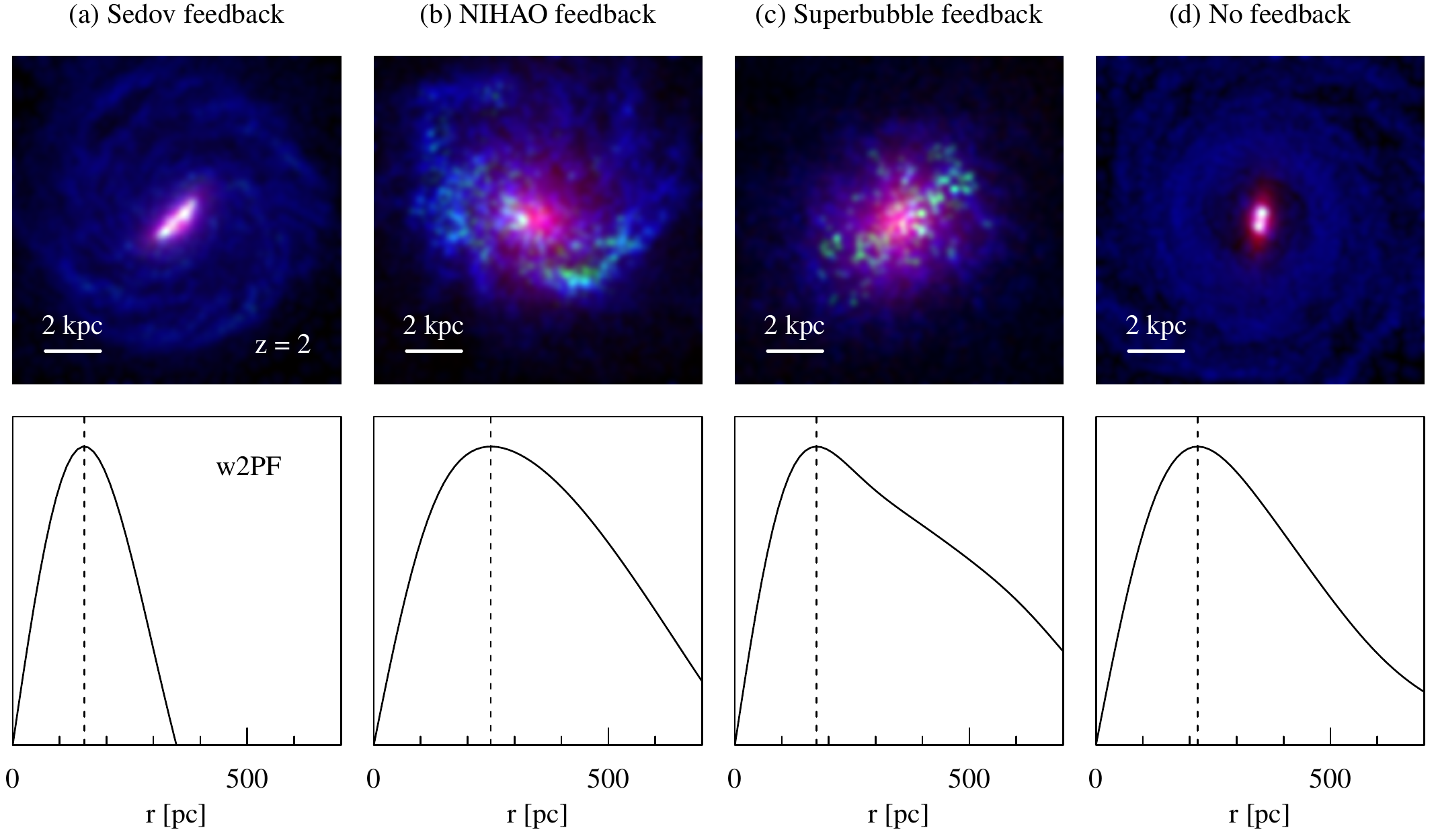}
\caption{Simulated Milky Way-like galaxy at $z=2$ with four different feedback models, described in Section~\ref{subsec:fb}. Upper row: false-color face-on galaxy images, showing the stellar surface density (red), star formation rate surface density (green) and cold gas (blue). Bottom row: weighted 2-point correlation functions as a function of scale $r$. Peak values $r_{\rm peak}$ values are shown as dashed lines.}
\label{fig:sim_galaxies}
\end{figure*}

\section{Clump-scalings in simulated galaxies}\label{sec:simulations}

The theoretical scaling relation for marginally stable ($Q\approx1$) disks shown as grey region in Figure \ref{fig:all_points} relies on a simple calculation that neglects local asymmetries, complex accretion dynamics and stellar feedback. To test whether the relation still applies in the presence of more complex processes, we now consider a zoom-simulation of a galaxy at $z=2$, near its peak star-formation, in a cosmological context. 

The simulated galaxy is a main-sequence object that ends up in a $10^{12}\rm M_{\odot}$ halo at $z=0$. This is likely less massive than the descendants of clumpy systems typically studied at higher redshifts, however we do not expect this difference to affect the physics analysed in this study. The galaxy is simulated four times using four different feedback models, including a no-feedback model. All four runs are re-simulations of the galaxy g8.26e11 from the NIHAO simulation project using identical initial conditions and the same cosmological environment (see \citealp{Wang2015MNRAS.454...83W}). The new runs use the updated hydrodynamics code \texttt{GASOLINE2} \citep{Wadsley2017MNRAS.471.2357W}, which improves on the original NIHAO galaxy that used \texttt{GASOLINE} \citep{Wadsley2004NewA....9..137W}. The runs use a flat $\Lambda$CDM cosmology with \citet{Planck2014A&A...571A..16P} parameters: Hubble parameter $H_{0}=67.1\ {\rm km\ s^{-1}}$, matter density $\Omega_{\rm M}=0.3175$, dark energy density $\Omega_{\rm \Lambda}=0.6824$, radiation density $\Omega_{\rm r}=0.00008$, baryon density $\Omega_{\rm b}=0.0490$, power spectrum normalization $\sigma_{8} = 0.8344$ and power spectrum slope $n_{\rm s} = 0.9624$.

Simulated with the standard NIHAO feedback, this galaxy has global properties similar to those of the Milky Way, with a final dynamical mass at $z=0$ of about $10^{12} M_{\odot}$. At $z=2$, this galaxy has a dynamical mass of $5\cdot10^{11} M_{\odot}$ and a stellar+cold gas mass of $5\cdot10^{10} M_{\odot}$ to $10^{11} M_{\odot}$, depending on the feedback model. The galaxy is simulated with $10^{6}$ dark matter+baryonic particles. The different feedback models and their results are discussed in the following section.

\subsection{Feedback models}\label{subsec:fb}

The first feedback model we consider is purely thermal resulting from the blastwave of stars within the mass range of $8 M_{\odot}< M_{\rm star} < 40 M_{\odot} $ undergoing a core-collapse supernova (SN) explosion. This \emph{Sedov Feedback} is implemented using the formalism of \citet{Stinson2006MNRAS.373.1074S}, which ensures that the energy and metals are ejected by the wave with cooling turned off for the particles within the blast radius. However, the high-density gas in the vicinity of the blast radius is allowed to cool and generally radiates the energy away efficiently. The resulting galaxy, shown in Figure \ref{fig:sim_galaxies} (a), therefore exhibits a lack of star formation in the outer regions.

The fiducial model used in NIHAO simulations employs the Early Stellar Feedback (ESF) model, originally explored by \citet{Stinson2013MNRAS.428..129S}, in addition to the \emph{Sedov Feedback} model. The ESF model incorporates the feedback mechanism due to the radiation of pre-SN massive stellar population, which adds a pathway for young massive stars to provide an ionizing source to the surrounding media and hence release energy into the ISM. Typically, an O-type star releases about $\sim 2\times10^{50}$ erg of energy per $M_{\odot}$ during the few $\rm Myrs$ between formation and SN explosion. This is comparable to the energy released by the SN itself. The fraction of the flux emitted in the ionizing UV was taken as $10\%$ by \citet{Stinson2013MNRAS.428..129S}. However, in our study we increase this stellar feedback efficiency to $\epsilon_{ESF} = 13\%$ to ensure better agreement with the mean stellar-to-halo mass relation derived from abundance matching \citep{Behroozi2013ApJ...770...57B}. Radiative cooling is allowed in this process. Figure \ref{fig:sim_galaxies} (b) shows the result of the NIHAO feedback model with increased star formation in the outskirts leading to a more realistic galaxy.


The third feedback scheme used in our analysis treats the evolution of clustered young stellar population as \emph{Superbubbles} wherein the associated structure is multi-phased with the feedback energy in hot phase being thermal while the cold expanding shell contains kinetic energy. Numerical simulations of the early stages of the superbubble are resolution-dependent and therefore computed using the analytical formalism of \citet{Keller2014MNRAS.442.3013K}, which employs thermal conduction to smoothly transition between the phases and hence provide a resolution insensitive result. We modify the \texttt{GASOLINE} code as per their suggestions to implement this technique in the zoom-in isolated galaxy simulation. Figure \ref{fig:sim_galaxies} (c) shows that this feedback model enhances star formation in the outer regions of the galaxy. 

Finally, we also evolve the galaxy in the absence of a feedback mechanism (Figure \ref{fig:sim_galaxies} (d)) to quantify the extent to which turning on a feedback mechanisms affects the clump physics.

\subsection{Results}

The galaxy simulations evolved using the four feedback models exhibit vastly different morphologies as shown in Figure \ref{fig:sim_galaxies}. In particular, in the absence of feedback, we run into the classical `angular momentum catastrophe' where the stellar disk (red) is too small and bulgy. Interestingly, the no-feedback model still produces an extended cold gas disk (blue), but its mass is negligible compared to the stellar mass of this galaxy ($9\%$), as well as compared to the cold gas of the other galaxies ($\sim20\%$).


To measure the clump-sizes of these galaxies, we treat them in a similar way to the observations: Each galaxy is projected (face-on) onto a two-dimensional grid with $700\times700$ cells. As detailed in Section~2, stellar particles younger than $10~\rm Myr$ are taken to represent the star-formation rate surface density (green channel in Figure~\ref{fig:sim_galaxies}), while  all stellar particles are used for the normalising global stellar surface density (red channel). For reference, the cold gas ($T<10^4\rm~K$) is shown in the blue channel; hence regions where all three components are abundant appear white. No radiative transfer is accounted for in producing the images. The clump size is then determined using the w2PF of Section~2. The clump size does not depend on the number of grid cells (resolution) as long as the cells are smaller than the clumps. The only source of uncertainty applied to $r_{\rm clump}$ is due to \emph{Sample variance}.

To estimate velocity dispersion $\sigma$ we use the standard deviation of the line-of-sight (vertical) velocity of gas particles as would be observed in a real observation. We find this dispersion to deviate no more than 30\% from the radial velocity dispersion in all the runs. The maximum rotation velocity $V$ and the half mass radius are straightforward to compute and have negligible errors. 

Figure \ref{fig:all_points} shows that the scaling relation of F17 holds for all feedback models. However, the galaxy can move along this relation depending on the feedback. The largest jump occurs in the case of no feedback, where $r_{\rm clump}/r_{\rm disk}$ increases by a factor of $\sim 2$. This is mostly because of the disk being too small, however the other parameters compensate this change, such the galaxy falls onto the F17 scaling relation, i.e.\ back onto the Jeans' length prediction. As the intensity of feedback increases the galaxy resembles the observed turbulent disks in morphology and lies close to the DYNAMO-HST sample. This indicates that the simple scaling model of F17 is a useful tool to diagnose \emph{in-situ} clump formation via VDIs.

\section{Discussion and Conclusion}\label{sec:discussion}

In this paper we used the two-point statistics of A17 to estimate the characteristic size of star-forming clumps in the DYNAMO-HST galaxies and isolated galaxy simulations with four feedback models. We found that the estimated clump sizes are in good agreement with the previous study of F17, which uses a more subjective clump-by-clump analysis to infer an average clump size (which would result in very different sizes if higher-resolution images were available). It follows that the updated clump sizes (measured using the two-point statistics) remain on the scaling relation displayed in Figure~\ref{fig:all_points}. This scaling relation is therefore robust under a more objective clump size determination, which would remain constant under increasing spatial resolution revealing increasing levels of fragmented substructure.

Secondly, using a zoom-in simulation of a single Milky Way-like galaxy with four different feedback models, we verified that the clump size scaling relation of F17 remains valid in the presence of realistic galaxy formation physics. Interestingly, the relation holds regardless of the feedback model. This finding aligns with the results of \citet{Hopkins2012MNRAS.427..968H}, who show that at low redshift, in the absence of mergers, the global Toomre parameter of isolated Milky Way-type galaxies is self-regulated and independent of the underlying microphysics.

An important ramification of the simulations presented in Figure~\ref{fig:sim_galaxies} is that, while all feedback models satisfy the basic Toomre model visualized in Figure~\ref{fig:all_points}, the supra-clump structure (e.g.~total number of clumps, their physical sizes and spatial extent) of these galaxies depends enormously on the feedback model. A more in-depth analysis of how these properties depend explicitly on the radiation pressure \citep{Mandelker2017MNRAS.464..635M} drew comparable conclusions. Similarly, a direct comparison of the global clump patterns produced by blastwave (Sedov) versus Superbubble feedback \citep{Mayer2016ApJ...830L..13M} predicts easily observable differences between these models in the macroscopic distribution of clumps.

Returning to the interesting finding that the $r_{\rm clump}/r_{\rm disk}$--$\sigma/V$ relation is almost universal without a strong dependence on the feedback model, we caution that this result is only based on simulations of a single halo. It would be interesting to vastly expand these simulations to cover a wide parameter space, especially a larger range of halo masses, merger scenarios and redshifts.

Overall, this work emphasizes the usefulness of the w2PF (A17) to measure clump sizes in observed and simulated datasets and it demonstrates the power of the clump size scaling relation of F17 to diagnose \emph{in-situ} clump formation via VDIs. This parallels recent developments on spatial correlations of star-forming disks on scales larger than clumps \citep{Combes2012A&A...539A..67C,Hopkins2012MNRAS.423.2016H,Grasha2017}, as well as within individual clumps \citep{Guszejnov2016MNRAS.459....9G}. Spatial correlations can therefore be regarded as an essential modern tool for studying the physics of star-forming disks.

\section*{Acknowledgements}
LW thanks Ben Keller for his help in setting up the Superbubble feedback model. The simulations were supported by high performance computing project pawsey0201 and was carried out on Magnus cluster at the Pawsey computing centre at Perth. DBF, DO and KG acknowledge support from Australian Research Council grants DP130101460 and DP160102235. DBF acknowledges support from an Australian Research Council Future Fellowship
(FT170100376) funded by the Australian Government. The Hubble Space Telescope data in this program are drawn from the HST program PID 12977 (PI Damjanov). DO thanks Chris Power for insightful discussions.

\bibliography{draft.bbl}
\bibliographystyle{apj}

\begin{appendix}

\begin{table*}[h]
\def\arraystretch{1.5}
\centering
\begin{tabular}{|C{1.75cm}|C{2.5cm} | C{2.5cm} | C{2.5cm} |C{2.5cm} |C{3.5cm} |}
\hline
 Galaxy & $z$ & $2r_{\rm clump}$ (pc) & $2r_{\rm disk}$ (kpc)& $\sigma/V$& Morphology\\  
 \hline
 G04-1 	& 0.1323	& $590^{+118}_{\rm -124} $ 		& $2.75$ 	& $0.19\pm0.09$	& Turbulent disk \\  
 G20-2 	& 0.1411	& $614^{+123}_{\rm -129} $ 		& $2.1$	& $0.49\pm0.07$	& Turbulent disk \\   
 D13-5 	& 0.0753	& $470^{+94}_{\rm -99}$ 		& $2.04$	& $0.24\pm0.02$	& Turbulent disk \\ 
 G08-5 	& 0.13217	& $462^{+92}_{\rm -97} $ 		& $1.84$	& $0.26\pm0.07$	& Turbulent disk \\  
 D15-3 	& 0.06712	& $196^{+44}_{\rm -196} $ 		& $2.2$	& $0.19\pm0.02$	& Turbulent disk (unresolved clumps)\\  
 G14-1 	& 0.13233	& $424^{+85}_{\rm -89} $ 		& $1.12$	& $0.51\pm0.07$	& Turbulent disk \\  
 C13-1 	& 0.07876	& $346^{+78}_{\rm -346} $ 	& $4.21$	& $0.13\pm0.04$	& Turbulent disk (unresolved clumps)\\  
 A04-3 	& 0.06907	& $150^{+34}_{\rm -150} $ 		& $3.58$	& $0.05\pm0.03$	& Normal spiral galaxy (unresolved GMCs)\\  
 H10-2 	& 0.14907	& $602^{+120}_{\rm -126} $ 		& $2.55$	& $0.95\pm0.34$	& Major Merger \\  
 G13-1 	& 0.13876	& $658^{+132}_{\rm -138} $ 		& $2.59$	& $0.68\pm0.03$	& Major Merger \\  
 \hline
 NGC 5194	& 0 (8 Mpc)	& $58\pm12$ 		& $5.6$	& $0.05\pm0.01$	& Normal spiral galaxy \\ 
 \hline
 Sedov Feedback 	& 2	& $220\pm44 $ 		& $1.96$	& $0.27\pm0.05$	& Turbulent disk \\  
 NIHAO Feedback 	& 2	& $460\pm92 $ 		& $3.74$	& $0.38\pm0.08$	& Turbulent disk \\  
 Superbubble Feedback 	& 2	& $280\pm56 $ 		& $3.14$	& $0.26\pm0.05$	& Turbulent disk \\  
 No Feedback 	& 2	& $380\pm76 $ 		& $1.7$	& $0.56\pm0.11$	& Turbulent disk \\  
  \hline
 
 \hline
\end{tabular}
\caption{Observed and simulated galaxy properties. The values for the DYNAMO-HST observations come from F17 and those for NGC 5194 from \citet{Leroy2008AJ....136.2782L}. We list the \textit{double} of the Gaussian radius for clumps and effective radius for disks for consistency with the definition of F17.}
\label{table:params}
\end{table*}

\end{appendix}

\end{document}